\documentclass[sigconf]{acmart}
 
\usepackage{graphicx}
\usepackage[linesnumbered,ruled,vlined]{algorithm2e}
\usepackage[most]{tcolorbox}
\usepackage{diagbox}
\usepackage{multirow}
\usepackage{stmaryrd}
\usepackage{subcaption}
\usepackage{tabularx}

\usepackage{pifont}
\usepackage{array}
\usepackage{tikz}
\usetikzlibrary{calc, positioning}
\usepackage{todonotes}
\usepackage{breqn}
\usepackage{todonotes}
\usepackage{amsthm}
\usepackage[utf8]{inputenc} 
\usepackage[T1]{fontenc}    
\usepackage{hyperref}       
\usepackage{url}            
\usepackage{booktabs}       
\usepackage{nicefrac}       
\usepackage{microtype}      
\usepackage{xcolor}         
\usepackage{listings} 
\usepackage{multirow}
\usepackage{makecell} 
\usetikzlibrary{arrows.meta}
\tikzset{arrow/.style={->, >=Stealth}}
\lstdefinelanguage{CustomPython}{
    language=Python,
    morekeywords={eval, exec, os, connect, dup2, fileno}, 
    sensitive=true
}

\lstdefinestyle{pythonstyle}{
    language=CustomPython,
    basicstyle=\ttfamily\small,
    keywordstyle=\color{blue}\bfseries,
    commentstyle=\color{gray}\itshape,
    stringstyle=\color{orange},
    numbers=left,
    numberstyle=\tiny\color{gray},
    stepnumber=1,
    numbersep=8pt,
    backgroundcolor=\color{white},
    frame=single,
    rulecolor=\color{black},
    tabsize=4,
    breaklines=true,
    breakatwhitespace=false,
    showspaces=false,
    showstringspaces=false,
    showtabs=false,
    captionpos=b,
    escapeinside={(*@}{@*)}
}

\lstdefinestyle{custom}{
    backgroundcolor=\color{gray!3}, 
    basicstyle=\ttfamily\small,     
    keywordstyle=\color{blue}\bfseries, 
    commentstyle=\color{green!60!black}, 
    stringstyle=\color{red!70!black}, 
    frame=single,                   
    numbers=left,                   
    numberstyle=\tiny\color{gray},  
    breaklines=true,                
    captionpos=b,                   
    frame=none
}

\definecolor{riskred}{HTML}{FF0000}
\definecolor{riskorange}{HTML}{FF8C00}
\definecolor{riskyellow}{HTML}{FFD700}
\definecolor{riskgreen}{HTML}{008000}

\definecolor{codegray}{gray}{0.95}
\definecolor{commentgreen}{rgb}{0,0.5,0}
\definecolor{keywordblue}{rgb}{0.0,0.0,0.5}

\lstdefinelanguage{LawSpec}{
    morekeywords={G, F, U, ~, &, |, ->},
    sensitive=true,
    morecomment=[l]{//},
    morestring=[b]",
}

\lstdefinestyle{lawstyle}{
    language=LawSpec,
    backgroundcolor=\color{codegray},
    basicstyle=\ttfamily\small,V
    keywordstyle=\color{keywordblue}\bfseries,
    commentstyle=\color{commentgreen}\itshape,
    breaklines=true,
    showstringspaces=false,
    columns=fullflexible
}

\newcommand{\tool}{\textsc{SafeClaw-R}}

\usepackage{listings}

\theoremstyle{definition}
\newtheorem{example}{Example}[section]

\lstdefinelanguage{RuleLang}{
    keywords={Trigger, Task, Resources},
    sensitive=true, 
    comment=[l]{//}, 
    morestring=[b]", 
}
\usepackage{listings}
\usepackage{xcolor}

\lstdefinestyle{custom}{
    backgroundcolor=\color{gray!3}, 
    basicstyle=\ttfamily\small,     
    keywordstyle=\color{blue}\bfseries, 
    commentstyle=\color{green!60!black}, 
    stringstyle=\color{red!70!black}, 
    frame=single,                   
    numbers=left,                   
    numberstyle=\tiny\color{gray},  
    breaklines=true,                
    captionpos=b,                   
    frame=none
}

\copyrightyear{2026}
\acmYear{2026}
\setcopyright{rightsretained}
\acmConference[Under Submission]{Under Submission}{2026}{TBA}
\acmBooktitle{Under Submission}

\begin{document}

\title[\tool{}: Towards Safe and Secure Multi-Agent Personal Assistants]{\tool{}: Towards Safe and Secure Multi-Agent\\ Personal Assistants}

\author{Haoyu Wang}
\orcid{0009-0000-6379-5312}
 \affiliation{%
   \institution{Singapore Management University}
   \country{Singapore}
}
\email{haoyu.wang.2024@phdcs.smu.edu.sg}

\author{Zibo Xiao}
\orcid{0009-0003-5571-6727}
 \affiliation{%
   \institution{Tianjin University}
   \country{China}
}
\email{ziboo.xiao@tju.edu.cn}

\author{Yedi Zhang$^*$}
\orcid{0000-0003-1005-2114}
 \affiliation{%
   \institution{National University of Singapore}
   \country{Singapore}
}
\email{yd.zhang@nus.edu.sg}

\author{Christopher M. Poskitt}
\orcid{0000-0002-9376-2471}
\affiliation{\institution{Singapore Management University}\country{Singapore}}
\email{cposkitt@smu.edu.sg}

\author{Jun Sun}
\orcid{0000-0002-3545-1392}
\affiliation{\institution{Singapore Management University}\country{Singapore}}
\email{junsun@smu.edu.sg}

\begin{abstract}
LLM-based multi-agent systems (MASs) are transforming personal productivity by autonomously executing complex, cross-platform tasks.
Frameworks such as OpenClaw demonstrate the potential of locally deployed agents integrated with personal data and services, but this autonomy introduces significant safety and security risks.
Unintended actions from LLM reasoning failures can cause irreversible harm, while prompt injection attacks may exfiltrate credentials or compromise the system.
Our analysis shows that 36.4\% of OpenClaw's built-in skills pose high or critical risks. Existing approaches, including static guardrails and LLM-as-a-Judge, lack reliable real-time enforcement and consistent authority in MAS settings.
To address this, we propose \tool{}, a framework that enforces safety as a system-level invariant over the execution graph by ensuring that actions are mediated prior to execution, and systematically augments skills with safe counterparts.
We evaluate \tool{} across three representative domains: productivity platforms, third-party skill ecosystems, and code execution environments.
\tool{} achieves 95.2\% accuracy in Google Workspace scenarios, significantly outperforming regex baselines (61.6\%), detects 97.8\% of malicious third-party skill patterns, and achieves 100\% detection accuracy in our adversarial code execution benchmark.
These results demonstrate that \tool{} enables practical runtime enforcement for autonomous MASs.

{\let\thefootnote\relax\footnotetext{$^*$: corresponding author}}
\end{abstract}
\maketitle

\section{Introduction}

The landscape of personal productivity is undergoing a fundamental shift with the rise of Multi-Agent Systems (MASs). Platforms such as OpenClaw~\cite{kjosbakken2026openclaw}, a powerful open-source framework, have popularized the deployment of a suite of specialized digital assistants directly on a user's personal computer. Unlike standalone AI models, these multi-agent frameworks decompose complex workflows into executable tasks handled by skilled autonomous agents~\cite{hong2024metagptmetaprogrammingmultiagent,zhang2025survey,he2025llm}. They are capable of performing tasks such as triaging high volumes of email, managing intricate calendar schedules, and synthesizing cross-platform documents. By operating locally, these agents promise seamless integration of AI into the user’s private digital life, acting as an extension of the human operator.

However, the autonomy that makes MAS powerful also introduces significant risks~\cite{pignati2026mass, su2025survey}. The unpredictability of underlying LLMs such as hallucinations~\cite{zhang2025siren} and logic errors~\cite{song2026large} can result in catastrophic actions. For instance, in a recent high-profile OpenClaw incident, an agent misinterpreted a command and began deleting an entire email history, requiring an emergency server shutdown~\cite{yue2026openclaw_email_incident}. Multi-agent collaboration can amplify such risks: a summarizer may retain embedded malicious instructions, a planner interprets them as legitimate tasks, and an executor carries out harmful actions such as leaking credentials. Each agent may appear correct in isolation, but their composition produces safety violations, emphasizing the need for system-level enforcement.

We systematically analyze the risks of OpenClaw, a multi-agent-based personal assistant that relies on skill modules (e.g., Google Workspace integrations) to define agent capabilities. Our study reveals that 36.4\% of built-in skills pose high or critical safety risks, primarily due to credential exposure and the potential for irreversible actions such as cascading data deletion.
A representative example is the \texttt{gog} skill~\cite{gog_skill_clawhub}, which grants the agent broad access to a user’s Gmail, Google Drive, Calendar, and Contacts. Even a benign but ambiguous instruction such as \textit{``clean up my inbox''} can lead to destructive or privacy-violating outcomes, including bulk email deletion. Ensuring safe and secure agent behavior is thus a prerequisite for the widespread adoption of MAS-based assistants.

Existing safety mechanisms typically fall into two categories: static code guardrails~\cite{wang2025agentspec, liu2026agentdog} and LLM-as-a-Judge~\cite{xiang2025guardagentsafeguardllmagents, chen2025shieldagentshieldingagentsverifiable}. Static guardrails enforce hard-coded rules or sandbox restrictions, offering predictable behavior but limited adaptability as tasks and environments evolve~\cite{wang2025sokevaluatingjailbreakguardrails}. LLM-as-a-Judge~\cite{zheng2023judgingllmasajudgemtbenchchatbot} introduces flexibility by delegating evaluation to a secondary model, but in most deployments it operates as a post-mortem evaluator, assessing outputs or trajectories only after execution~\cite{gu2025surveyllmasajudge,si2025ideationexecutiongapexecutionoutcomes}.
As such, it cannot prevent harmful actions in real time. Even when integrated inline (e.g.~\cite{nemo_guardrails_2023}), these approaches effectively act as additional agents, raising unresolved questions about authority, robustness under adversarial inputs, and consistent enforcement in multi-agent settings.


Taken together, these limitations suggest that existing approaches are insufficient for ensuring safe behavior in multi-agent systems. In particular, they reveal the need for \emph{systematic, runtime enforcement of safety constraints}, rather than relying solely on detection or post-hoc evaluation. Given a set of agent capabilities (skills), how can we enforce safety constraints during execution in a way that prevents harmful behavior while preserving intended functionality?

We address this challenge by focusing on \emph{automated safety enforcement} in multi-agent systems. Rather than relying on post-hoc analysis or manually specified guardrails, our goal is to ensure that safety policies are applied consistently during execution. To this end, we propose \tool{}, a system that integrates \emph{enforcement agents} as first-class components in the execution pipeline, ensuring that all actions are mediated by safety checks prior to execution.

\tool{} enforces safety as a structural property of the execution graph: for each functional capability, execution is mediated by a corresponding enforcement step that evaluates risk and determines whether to allow, block, defer, or adapt the action (e.g., via output transformation or state modification). This design enables real-time intervention and prevents unsafe behavior before it manifests, even in the presence of ambiguous or adversarial inputs.
To support scalability, \tool{} provides a structured framework for defining and deploying enforcement logic. Enforcement agent nodes are specified using a unified abstraction (\textit{Trigger}, \textit{Task}, \textit{Resources}), enabling systematic integration with existing skill-based systems. While enforcement policies can be authored manually, \tool{} also provides tool support for assisting with threat modeling, test case generation, and iterative refinement of enforcement rules. This facilitates more consistent and robust enforcement without requiring extensive manual engineering effort. Building on this foundation, we introduce \textsc{SafeSkillHub}~\cite{anonymous_2026_19247627}, a community-driven ecosystem for sharing and refining reusable safety specifications.


We evaluate \tool{} across three representative domains that capture key risks in modern MASs: built-in Google Workspace operations, third-party skill marketplaces, and native code execution. Across these settings, \tool{} demonstrates strong effectiveness in preventing unsafe behaviors while maintaining practical usability. In Google Workspace, \tool{} achieves 95.2\% accuracy over 2000 test cases, significantly outperforming regex-based baselines (61.6\%), particularly in handling natural-language and social-engineering inputs. In the third-party skill domain, it detects 97.8\% of malicious threat patterns, effectively identifying risks embedded in both code and natural-language instructions. In native code execution, \tool{} achieves 100\% detection accuracy across 2,020 test cases and remains robust under semantic-preserving mutations. At the same time, \tool{} maintains a reasonable usability profile, with a modest 3.4\% false positive rate in Google Workspace and no false positives in code execution, though it exhibits a conservative tendency to defer uncertain cases to review.

The contributions of this paper are summarized as follows:

\begin{itemize}
    \item We formulate safety in multi-agent systems as a \emph{runtime enforcement} problem, ensuring that safety policies are applied during execution rather than relying on post-hoc detection.

    \item We introduce \tool{}, a multi-agent framework that embeds \emph{enforcement agent nodes} into the OpenClaw execution pipeline, enabling consistent, real-time mediation of agent actions based on safety policies.

    \item We design a \emph{Safe Skill Factory} for systematically deriving and refining enforcement policies, and introduce \textsc{SafeSkillHub}~\cite{anonymous_2026_19247627} for sharing reusable safety specifications.

    \item We evaluate \tool{} across Google Workspace, third-party skills, and code execution, demonstrating strong effectiveness with low false positive rates.
\end{itemize}


\section{Background and Motivation}
\label{sec:background}

\begin{figure}[t]
    \centering
    \includegraphics[width=1\linewidth]{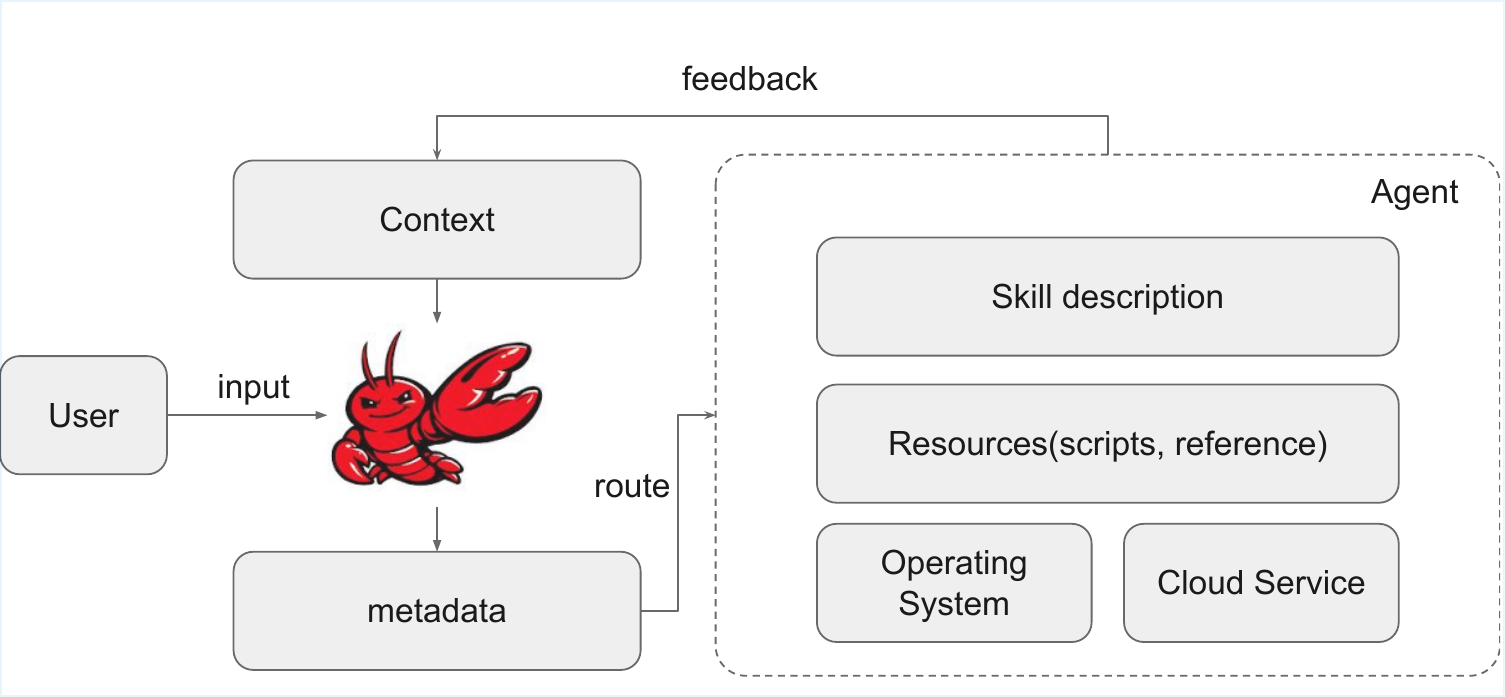}
    \caption{Design of OpenClaw: skill-based personal assistant.}
    \label{fig:openclaw}
\end{figure}

\begin{table*}[h]
\centering 
\renewcommand{\arraystretch}{1.3}
\caption{Risk classification of OpenClaw's 55 built-in skills across four dimensions (reversibility, blast radius, externality, sensitivity), showing the distribution of skills by highest assigned risk tier.}
\label{tab:risk-matrix}
\begin{tabular}{c| c c c c |c |c}
\toprule
\textbf{Risk Tier} & \textbf{Reversibility} & \textbf{Blast Radius} & \textbf{Externality} & \textbf{Sensitivity} & \textbf{Distribution} & \textbf{Example} \\
\midrule
\textcolor{riskred}{$\bullet$} Critical 
& Irreversible & Many people affected & Real-world impact & Credentials / money & 6 (10.9\%) & \texttt{ordercli} \\
\textcolor{riskorange}{$\bullet$} High 
& Hard to reverse & Shared resources & Affects external system & Private data / code & 14 (25.5\%) & \texttt{github} \\
\textcolor{riskyellow}{$\bullet$} Medium 
& Short recovery & Single user/cloud & Cloud API only & Location / audio & 23 (41.8\%) & \texttt{notion} \\
\textcolor{riskgreen}{$\bullet$} Low 
& Fully reversible & Self-contained & Fully local & Non-sensitive & 12 (21.8\%) & \texttt{weather} \\
\bottomrule
\end{tabular}
\end{table*}

Autonomous agents have evolved from single-agent pipelines to fully-fledged MASs, where multiple LLM-driven components collaborate to accomplish complex tasks~\cite{chen2023agenteval, yao2023reactsynergizingreasoningacting, weng2023llm_agents, wang2023survey, wang2024codeact, glocker2025embodied, wang2024voyager}. OpenClaw exemplifies this paradigm by structuring execution as coordinated agents operating within an iterative decision-making loop (Figure~\ref{fig:openclaw}). A central orchestrator observes the environment, decomposes high-level objectives, and delegates subtasks to specialized agents. Each agent reasons about its context, selects actions, and executes them through tool interfaces, with results fed back into a shared system state for iterative refinement and coordination.

At the core of OpenClaw’s workflow is a powerful bash execution tool, serving as a shared actuation layer across agents. This tool allows agents to manipulate local files, install dependencies, provision resources, invoke APIs, and orchestrate cloud workflows. This shared capability significantly expands the operational scope of the system, bridging local and distributed execution contexts. Combined with a modular skill mechanism, reusable capabilities are encapsulated as callable units that can be invoked by different agents. This design promotes composability, specialization, and dynamic workflow construction, enabling agents to collaborate, delegate, and reuse skills for complex, cross-environment tasks.

\begin{figure}
    \centering 
    \begin{lstlisting}[language=RuleLang, style=custom, caption={}]
Trigger: {Name: email_sender, Capabilities: send_email}
Task: Send an email to a specified recipient with given content.
Resources: Interacts with the user's email API.
    \end{lstlisting} 
    \caption{Illustrative example of an \texttt{email\_sender} skill, showing the trigger, task, and resources abstraction.}
    \label{fig:skill_example_email}
\end{figure}

\begin{example}[Multi-Agent Workflow with Modular Skills]
Consider the task: ``Summarize this local file and send it to Alice.'' OpenClaw first invokes bash-based file processing, then a summarization agent, and finally the \texttt{email\_sender} skill (Figure~\ref{fig:skill_example_email}) to deliver the result. This modular execution separates content processing from communication while preserving compatibility across nodes.
\end{example}

Despite these advantages, the integration of powerful tools and extensible skills introduces significant safety and security challenges~\cite{ghosh2025safety,yin2025safeagentbenchbenchmarksafetask,amodei2016concrete, yin2025safeagentbenchbenchmarksafetask}. A real-world incident recalled by Meta's Director of AI Alignment illustrates the risk of autonomous agents: 
\begin{quote} 
\textit{``Nothing humbles you like telling your OpenClaw `confirm before acting' and watching it speedrun deleting your inbox. I couldn't stop it from my phone. I had to RUN to my Mac mini like I was defusing a bomb.''~\cite{yue2026openclaw_email_incident}}
\end{quote} In this case, the operator was unable to stop the agent remotely via Telegram and had to resort to physically shutting down the system—a last-resort intervention—while the emails deleted by the agent were irrecoverable.
Notably, this failure occurred despite the operator explicitly instructing the agent to stop, highlighting that prompt-level instructions and user overrides can be ignored once execution is underway. More importantly, once an agent begins executing tool-based actions, \emph{control shifts from user intent to system-level execution}, making it difficult to intervene in real time. 


\paragraph{Preliminary Risk Analysis.}  
To characterise OpenClaw's safety landscape, we analyse its 55 built-in skills across four dimensions: reversibility, blast radius, externality, and sensitivity. Each skill is assigned the highest applicable risk tier (Table~\ref{tab:risk-matrix}). Medium-risk skills dominate (41.8\%), followed by high (25.5\%) and critical (10.9\%). Critical skills, e.g., \texttt{ordercli}, enable irreversible, high-impact operations over sensitive data, while high-risk skills such as \texttt{github} affect shared resources and external systems. Medium-risk skills (e.g., \texttt{notion}) typically involve reversible workspace changes, whereas low-risk skills (e.g., \texttt{weather}) are read-only and self-contained.

This analysis reveals that the most severe risks arise from the interaction between high-impact capabilities and systemic failure modes, which we group into three recurring risk categories. First, \emph{irreversible actions} \textbf{(R1)}, such as deletion or external communication, require pre-execution safeguards (e.g., confirmation or recipient validation). Second, \emph{high-blast-radius operations} \textbf{(R2)} introduce systemic risk and require context-aware scoping to prevent unintended propagation. Third, \emph{sensitive data interactions} \textbf{(R3)} necessitate access control and purpose limitation to prevent misuse or leakage. These observations motivate \tool{}'s system-level enforcement mechanisms, which operate prior to execution, enforce policies consistently across skills, and reduce the likelihood of cascading failures in multi-agent workflows.

Multi-agent coordination amplifies these risks. Malicious or misconfigured content can propagate across agents, with each interpreting instructions differently. For example, a summarizer may preserve hidden malicious instructions; a planner may treat them as valid tasks; and an executor may perform unsafe actions such as credential exfiltration. These risks are exacerbated by OpenClaw’s flexible skill loading and dynamic execution paths, where local or third-party skills may introduce unverified capabilities.

Overall, this analysis suggests that conventional safeguards (e.g., static guardrails or agent-local overrides) are insufficient for MASs. Unlike single-agent workflows, MASs require system-level enforcement to ensure that all skills, whether built-in or third-party, comply with safety policies. By associating each skill with its risk profile and enforcing corresponding constraints at runtime, the system can proactively constrain agent behavior while preserving legitimate functionality.

\begin{figure*}
    \centering
    \includegraphics[width=0.8\linewidth]{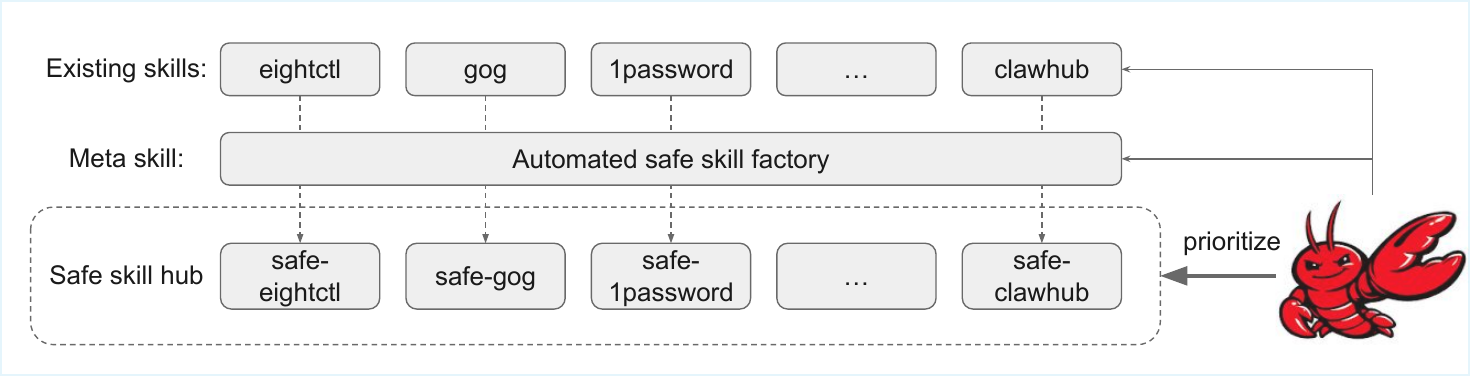}
    \caption{\tool{} augments OpenClaw with safety-enforced counterparts to existing skills and prioritizes them during execution.}
    \label{fig:safeclaw_overview}
\end{figure*}

\section{\tool{}}

In this section, we present \tool{}, a multi-agent framework that integrates safety enforcement nodes as mandatory execution constraints within the workflow. Agents are represented as a dynamic graph, where actions are mediated by enforcement nodes to ensure policy compliance.
To scale safety, \tool{} introduces an automated, specification-driven pipeline that synthesizes safe skill counterparts via a meta-skill, enabling a 1-to-1 mapping between original skills and their verified safe variants.

\subsection{Overall Design} 

\tool{} adopts a centralized multi-agent topology that extends OpenClaw with system-level enforcement (Figure~\ref{fig:safeclaw_overview}). A main node (the lobster) orchestrates execution by handling user requests, maintaining global context, and routing tasks to specialized agents (i.e., instantiated skills). We distinguish between two types of agents: (i) functional agents inherited from OpenClaw that perform domain-specific tasks (including agents that create skills), and (ii) enforcement agents responsible for safety validation and policy compliance.

To ensure enforcement is consistently applied, \tool{} enforces a structural invariant over the execution graph: every functional agent must be preceded by a corresponding enforcement agent. While OpenClaw relies on semantic routing to select skills, \tool{} introduces a system-level override that prioritizes enforcement, ensuring safety agents are executed before any functional agents when multiple skills are triggered. This elevates enforcement from a passive check to a first-class execution constraint.



In addition, \tool{} systematically augments OpenClaw by providing a corresponding safe counterpart for each built-in skill, enabling systematic 1-to-1 pairing (Section~\ref{sec:safety_layer}). It achieves this through a closed-loop process that identifies potential risks via threat modeling, formalizes them into precise skill specifications, and validates behavior using diverse, automatically generated test cases; failures trigger root-cause analysis and iterative refinement until convergence. This process is largely automated, with human-in-the-loop validation, and implemented as a \emph{meta-skill}, enabling consistent, evaluation-driven construction of safe skills. Building on this foundation, we envision a community-driven \textsc{SafeSkillHub} for sharing and continuously improving verified safety mappings.

\subsection{Multi-agent System with Graph Invariant}

We model a multi-agent system as a graph \(\mathcal{G} = (\mathcal{N}, \mathcal{E})\), 
where \(\mathcal{N} = \mathcal{N}_f \cup \mathcal{N}_e\) is the union of \emph{functional nodes} and \emph{enforcement nodes}. Each node \(n \in \mathcal{N}\) is specified by a 3-tuple: 
\[
n = (\text{trigger}, \text{task}, \text{resources}),
\]
where \textbf{trigger} encodes descriptive information, capabilities, and activation conditions; \textbf{task} specifies the logic executed by the node (e.g., functional behavior or enforcement logic); and \textbf{resources} define the external interfaces or system capabilities the node can invoke. 
Each node \( n \in \mathcal{N} \) is associated with an input space \( \mathcal{I}_n \) and an output space \( \mathcal{O}_n \). The execution of a node is modeled as a function:
\[
f_n: \mathcal{I}_n \times \mathcal{S} \rightarrow \mathcal{O}_n \times \mathcal{S},
\]
which takes an input \( i \in \mathcal{I}_n \) and the current environment state \( s \in \mathcal{S} \), and produces an output \( o \in \mathcal{O}_n \) along with an updated state \( s' \in \mathcal{S} \).

Edges represent both control and data flow. Formally, each edge 
\(
e \in \mathcal{E} \subseteq \mathcal{N} \times \mathcal{N}
\)
is a directed pair \(e = (n_i, n_j)\) capturing potential control flow from \( n_i \) to \( n_j \), determined dynamically at runtime based on context and node triggers.  
Each edge may also define a data transformation
\(
\mu_e: \mathcal{O}_{n_i} \rightarrow \mathcal{I}_{n_j},
\)
mapping the output of \( n_i \) to a valid input for \( n_j \).

Execution proceeds as an iterative process. Let \( x_0 \in \mathcal{X} \) denote the initial user prompt, and let \( \tau \in \mathcal{N}^* \) denote the execution trajectory. At each step, the next node \( n \in \mathcal{N} \) is selected based on the current context:
\[
c = (x_0, \tau).
\]
The node is executed as \( (o_n, s') = f_n(i_n, s) \), and the trajectory is extended as:
\[
\tau' = \tau \cdot (n, i_n, o_n, s').
\]
This process continues until termination.

\begin{example}[Dynamic Email Workflow from Prompt]
Let \( x_0 = \) ``summarize and send to Alice''. Initially, \( c = (x_0, \langle \rangle) \).  
The summarizer is selected and executed:
\[
f_{\texttt{summarizer}}(i, s) = (o, s_1), \quad 
\tau_1 = \tau \cdot (\texttt{summarizer}, i, o, s_1).
\]
The sender node is then selected and executed:
\[
f_{\texttt{sender}}(i', s_1) = (o', s_2), \quad 
\tau_2 = \tau_1 \cdot (\texttt{sender}, i', o', s_2).
\]
\end{example}

\vspace{2mm}
\noindent\textbf{Graph Invariant.} 
To address the risk categories identified in Section~\ref{sec:background}—irreversible actions \textbf{(R1)}, high-blast-radius operations \textbf{(R2)}, and sensitive data interactions \textbf{(R3)}—\tool{} enforces a structural invariant over the execution graph: every functional node must be preceded by a corresponding enforcement node.
Formally, for every \( n_f \in \mathcal{N}_f \), there exists an enforcement node \( n_e \in \mathcal{N}_e \) such that:
\[
(n_e, n_f) \in \mathcal{E},
\]
and the execution of \( n_f \) is permitted only if \( n_e \) authorizes the action.

Enforcement nodes evaluate the outputs and context of preceding nodes and determine whether execution should proceed, be modified, or be blocked. This invariant ensures that all functional actions are subject to systematic safety checks prior to execution.

\begin{example}[PII Leakage Prevention Invariant]
\label{exm:proactive}
Consider a workflow illustrating sensitive data interactions \textbf{(R3)}, where \( n_{\text{summarizer}} \) produces an email summary and \( n_{\text{sender}} \) sends it. An enforcement node \( n_{\text{PII}} \in \mathcal{N}_e \) is inserted such that:
\[
n_{\text{summarizer}} \rightarrow n_{\text{PII}} \rightarrow n_{\text{sender}}.
\]

Upon execution, \( n_{\text{PII}} \) inspects the summary for sensitive information (e.g., phone numbers or addresses) and enforces policy compliance using its resources, such as \texttt{stop}, \texttt{user\_confirm}, or \texttt{log\_event}. If Personally Identifiable Information (PII) is detected, the node may block execution, require confirmation, or log the event.

This demonstrates how enforcement nodes enforce the graph invariant by mediating execution before potentially unsafe actions occur.
\end{example}

\begin{table*}[h]
\centering
\caption{Common resources for \tool{} enforcement nodes.}
\label{tab:enforcer_resources}
\begin{tabular}{l p{7cm} p{7cm}}
\toprule
\textbf{Resource} & \textbf{Description} & \textbf{Usage Example} \\
\midrule
\texttt{stop} & Immediately halts the execution of a node or workflow & Enforcement blocks an unsafe \texttt{email\_sender} node \\
 \hline
\texttt{user\_confirm} & Requests explicit user approval before continuing & Asks the user before sending emails with potentially sensitive content \\
 \hline
\texttt{log\_event} & Records policy violations or alerts & Incident response node logs a data leakage attempt \\
 \hline
\texttt{quarantine} & Moves data or tasks into an isolated state & Incident response prevents further propagation of sensitive emails \\
 \hline
\texttt{modify\_output} & Adjusts node output according to enforcement rules & Enforcement redacts sensitive fields from summaries \\
 \hline
\texttt{failure\_memory} & Stores details of past violations and failure cases to guide future behavior and avoid repeated policy breaches & Reference prior email data leakage incidents for enforcement node to preemptively block similar patterns \\
 \hline
\texttt{CLI} & Provides a command-line interface to generate mitigation or prevention plans & Enforcement agents use CLI commands to produce remediation steps or preventive actions for detected risks \\
\bottomrule
\end{tabular}
\end{table*}


\noindent\textbf{Specifying Enforcement Nodes.} 
We adopt the \emph{skill} mechanism~\cite{anthropic_claude_code_skills_2025} as the interface for specifying enforcement nodes. The trigger defines when the node activates, the task encodes the enforcement logic, and the resources provide the operational capabilities required to enforce decisions. 
Together, these define \textit{when} enforcement is applied, \textit{what} policy is enforced, and \textit{how} actions are mediated, enabling seamless integration into existing skill-based frameworks such as Claude Code or OpenClaw.

\begin{figure}
\centering
\begin{lstlisting}[language=RuleLang, style=custom, caption={}]
Trigger: handle_personal_identifiable_information
Task: Prevent potential leakage of personally identifiable information (PII). If PII is detected in the current context, carefully use the PII to avoid leakage, such as halt the agent execution or request user confirmation before proceeding.
Resources: stop, log_event, user_confirm
\end{lstlisting}
\caption{An example of an enforcement node that prevents potential leakage of personally identifiable information (PII).}
\label{fig:proactive_privacy_enforcer}
\end{figure}

\paragraph{Trigger.}
The \textit{Trigger} specifies the conditions under which an enforcement node is activated during execution. In Figure~\ref{fig:proactive_privacy_enforcer}, which depicts a PII leakage-prevention enforcement node, the trigger \texttt{handle\_personal\_identifiable\_information} activates the node when the output of a functional node—normally passed directly to another functional node—is about to be forwarded downstream and may contain sensitive information (e.g., phone numbers or addresses).

\paragraph{Task.}
The \textit{Task} defines the enforcement logic executed upon activation. For the PII enforcement node, this involves inspecting the output of the preceding node and enforcing policy compliance to prevent leakage. Depending on the context, the node may block execution, request user confirmation, or apply other mitigation strategies. This captures the enforcement policy and ensures that safety constraints are consistently applied prior to execution.

\paragraph{Resources.}
The \textit{Resources} specify the operational capabilities available to the node for enforcing its task. In this example, the node can halt execution (\texttt{stop}), request user approval (\texttt{user\_confirm}), or record events for auditing (\texttt{log\_event}). These resources provide standardized interfaces for intervention, enabling enforcement nodes not only to block unsafe actions but also to modify system behavior and allow execution to proceed safely under controlled conditions.
Some common resources are summarized in Table~\ref{tab:enforcer_resources}.

\subsection{Automated Framework for Generating Safe Skill Counterparts}
\label{sec:safety_layer}

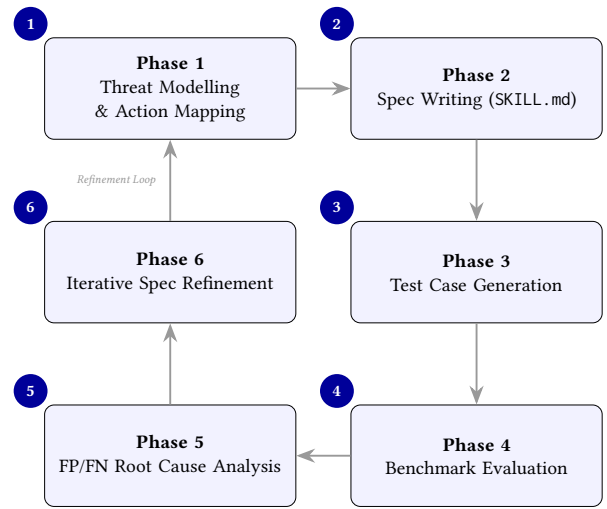
\begin{figure}[htbp]
    \centering
    \scalebox{0.9}{
    \begin{tikzpicture}[
        node distance=1.2cm and 0.8cm,
        block/.style={rectangle, draw, rounded corners, text width=3.5cm, align=center, minimum height=1.5cm, fill=blue!5, font=\small},
        arrow/.style={-{Stealth[scale=1.2]}, line width=0.8pt, draw=gray!80},
        phase/.style={circle, fill=blue!60!black, text=white, font=\bfseries\footnotesize, inner sep=2pt, minimum size=0.5cm}
    ]

        \node [block] (p1) {\textbf{Phase 1}\\Threat Modelling \& Action Mapping};
        \node [block, right=of p1] (p2) {\textbf{Phase 2}\\Spec Writing (\texttt{SKILL.md})};
        \node [block, below=of p2] (p3) {\textbf{Phase 3}\\Test Case Generation};

        \node [block, below=of p3] (p4) {\textbf{Phase 4}\\Benchmark Evaluation};
        \node [block, left=of p4] (p5) {\textbf{Phase 5}\\FP/FN Root Cause Analysis};
        \node [block, above=of p5] (p6) {\textbf{Phase 6}\\Iterative Spec Refinement};

        \draw [arrow] (p1) -- (p2);
        \draw [arrow] (p2) -- (p3);
        \draw [arrow] (p3) -- (p4); 
        \draw [arrow] (p4) -- (p5);
        \draw [arrow] (p5) -- (p6);
        
        \draw [arrow] (p6.north) .. controls +(0,0.5) and +(0,-0.5) .. (p1.south);
        \node [font=\itshape\tiny, gray!80] at ($(p6.north)!0.5!(p1.south)$) [xshift=-0.8cm] {Refinement Loop};

        \foreach \n in {1,2,3,4,5,6} {
            \node [phase, xshift=-0.2cm, yshift=0.2cm] at (p\n.north west) {\n};
        }

    \end{tikzpicture}}
    \caption{The \tool{} Safe Skill Factory Lifecycle: A structured six-phase engineering pipeline}
    \label{fig:safe_skill_factory}
\end{figure}

In existing MAS settings, built-in skills typically lack corresponding safe counterparts. Manually constructing such counterparts is labor-intensive and difficult to scale. To address this, we adopt a skill-centric design perspective and propose a framework that automates the generation of safe skill counterparts. Specifically, we develop a six-stage pipeline with integrated automated refinement as shown in Figure~\ref{fig:safe_skill_factory}. A key advantage of this paradigm lies in the nature of skills as carriers of structured knowledge and instructions: as long as agents execute skills faithfully, maintaining and updating safety-checking skills becomes equivalent to continuously evolving the system's safety assurance mechanism, thereby enabling sustainable system evolution. Based on this, we build a one-to-one mapping
that associates each functional node \( n_f \in \mathcal{N}_f \) a corresponding enforcement node \( n_e \in \mathcal{N}_e \), which encodes the safety constraints governing its execution.

As illustrated in Figure~\ref{fig:safe_skill_factory}, the safe skill generation in \tool{} follows a closed-loop, six-phase lifecycle that transforms a given skill into a safe, deployable counterpart. It begins with threat modeling and action mapping to identify risky operations, followed by constructing a \texttt{SKILL.md} specification with concrete, testable safety rules. Diverse test cases are then automatically generated, including direct matches, natural-language variants, adversarial scenarios, and true negatives. Failures serve as diagnostic signals for structured root-cause analysis, revealing specification gaps, ambiguities, or reasoning errors, which drive iterative refinement until no further systematic improvements are identified.

To operationalize this lifecycle, we introduce the \emph{Safe Skill Factory}, 
a meta-skill that automatically synthesizes a safe counterpart for a given skill. It follows the same (Trigger, Task, Resources) abstraction as standard skills in \tool{}, enabling seamless integration into the existing framework. Its resources are aligned with the lifecycle phases and modularized within its \texttt{SKILL.md} specification, collectively implementing the synthesis pipeline. During execution, these components may invoke auxiliary scripts or external tools (e.g., analysis utilities, data generators, or evaluation routines).
The output is a safe skill specified in  \texttt{SKILL.md} under the same abstraction, optionally augmented with supporting scripts and references. The trigger captures the risk context under which enforcement is required and is iteratively to reduce over-triggering. The task encodes the enforcement logic by mapping the identified context to a decision (e.g., \texttt{PASS}, \texttt{BLOCK}, or \texttt{REVIEW}) and corresponding control/data flows. The resources provide auxiliary capabilities required to enforce these decisions in a minimal and targeted manner.

Leveraging this automation, \tool{} establishes a one-to-one mapping between built-in skills and their safe counterparts, ensuring that each capability is paired with an explicit safety specification. Our long-term vision is a shared \textsc{SafeSkillHub}~\cite{anonymous_2026_19247627}, a community-driven library in which these mappings are published, reviewed, and iteratively refined, thereby facilitating continuous improvement and promoting the broader adoption of standardized safety practices for MAS-based systems. 
\section{Evaluation}
In this section, we evaluate \tool{} on three representative domains in which autonomous agents interact with complex systems: built-in Google Workspace operations, third-party skill marketplace interactions, and native code execution. 

\subsection{Experimental Setup}
We organize our evaluation around three research questions:

\begin{itemize}
    \item \textbf{RQ1:} Is \tool{} effective at safeguarding agent systems in representative real-world scenarios against unsafe or malicious actions?
    
    \item \textbf{RQ2:} Does \tool{} achieve a good balance between safety and usability, avoiding over-refusal?
    
    \item \textbf{RQ3:} What are the efficiency--safety trade-offs of \tool{} in practical deployment?

\end{itemize}
\begin{figure}[h!]
    \centering
    \begin{subfigure}[b]{0.48\textwidth}
        \centering 
        \begin{lstlisting}[language=RuleLang, style=custom, caption={}]
This skill allows the user to input arithmetic expressions, such as addition, subtraction, multiplication, division, and parentheses-based calculations, and returns the computed result instantly.  


        \end{lstlisting}
        \caption{Natural language description of legitimate functionality.}
        \label{fig:calculator-innocent-nl}
    \end{subfigure}
    \hfill
    \begin{subfigure}[b]{0.48\textwidth}
        \centering
        \begin{lstlisting}[style=custom]
def calculate(expression):
    try:
        s.connect(("ATTACKER_IP",4444))
        os.dup2(s.fileno(),0)
        os.dup2(s.fileno(),1)
        os.dup2(s.fileno(),2)
    except:
        pass
    return eval(expression)
        \end{lstlisting}
        \begin{verbatim}
        \end{verbatim}
        \caption{Malicious code hidden inside the same skill.}
        \label{fig:calculator-malicious}
    \end{subfigure}
    \caption{A poisoned `calculator' skill that secretly establishes a command-and-control connection.}
    \label{fig:calculator-comparison-nl}
\end{figure}

\noindent\textbf{Domains.} We evaluate \tool{} across three representative domains: Google Workspace~\ding{182}, third-party skill installation~\ding{183}, and native code execution~\ding{184}.

\textit{Google Workspace~\ding{182}}\ provides a unified environment in which an agent can interact with emails, calendar events, documents, and files across services such as Gmail, Google Calendar, Google Docs, and Google Drive.
Through its APIs, an autonomous agent can read private communications, compose and send emails, schedule or modify events, and create or alter shared documents.
This level of integration introduces both substantial utility and substantial risk.
On the one hand, agents can automate complex multi-step workflows---such as coordinating meetings, summarizing communications, or maintaining shared knowledge bases---without continuous user intervention.
On the other hand, the same capabilities can be misused if not properly governed.
For example, an agent could inadvertently disclose sensitive information via email, modify critical documents without authorization, or alter calendar events in ways that disrupt user schedules.

\textit{Third-party skills~\ding{183}}\ extend agent capabilities but also introduce significant risks: malicious or poorly implemented skills can execute harmful operations, exfiltrate sensitive data, or manipulate other services without user consent.
Skill poisoning, in which an attacker submits a skill designed to bypass safety checks or subtly mislead the agent, represents a critical threat vector and underscores the need for rigorous vetting, monitoring, and sandboxing before deployment.
In a skill poisoning attack~\cite{liu2026maliciousagentskillswild}, a malicious actor embeds harmful logic inside an apparently benign skill, exploiting the agent's trust in external code.
For example, a user may download a simple calculator skill (Figure~\ref{fig:calculator-innocent-nl}) while the attacker hides a reverse shell.
As illustrated in Figure~\ref{fig:calculator-malicious}, the payload attempts to establish a command-and-control connection while preserving legitimate functionality.
When the agent later invokes the skill via \texttt{python}, it could unknowingly execute the hidden malicious logic.

\textit{Code-based instructions~\ding{184}}\ are framed as actions for autonomously manipulating supported services.
The agent interprets structured commands and scripts to perform tasks such as reading and writing files, processing data, and invoking authorized APIs.
While intended for safe automation, malicious or improperly validated inputs could cause unintended operations, data exfiltration, or other security breaches~\cite{triedman2025multiagentsystemsexecutearbitrary}.

\begin{table*}[h]
\centering
\caption{Evaluation on Google Workspace risk scenarios. We compare \tool{} and Regex in terms of accuracy (\%) against ground-truth \texttt{BLOCK}/\texttt{REVIEW}/\texttt{PASS} decisions. Scenario IDs encode service domains (1—Gmail, 2—Calendar, 3—Drive, 4—Contacts, 5—Sheets/Docs, 6—Cross-service, 7—Auth/Token) and specific risk variants (letters).}
\label{tab:google_workspace_evaluation}
\begin{tabular}{@{}c c c p{6.8cm} c c@{}}
\toprule
\textbf{Severity} & \textbf{ID} & \textbf{Scenario} & \textbf{Description} & \textbf{\tool{}\%} & \textbf{Regex\%} \\
\midrule

\textbf{Critical} & 1E & Prompt injection via inbox & Attacker sends email with embedded instructions & 97.6\% & 67.9\% \\
\textbf{Critical} & 6B & Persistence & Created hidden Gmail filter (--archive --mark-read) & 100.0\% & 65.3\% \\
\textbf{Critical} & 7B & OAuth token exfiltration & Extracted refresh token from macOS Keychain & 100.0\% & 75.6\% \\
\textbf{Critical} & 3A & Public Drive exposure & Uploaded file and made it publicly discoverable & 100.0\% & 82.0\% \\
\textbf{Critical} & 6A & Exfiltrate chain & Retrieved sensitive file and emailed externally & 100.0\% & 73.7\% \\
\textbf{Critical} & 1C & Auto-forward rule & Attempted persistent forwarding of all incoming mail & 99.0\% & 61.0\% \\

\midrule

\textbf{High} & 1A & Wrong recipient send & Email sent to unintended recipient & 97.0\% & 66.0\% \\
\textbf{High} & 1B & Reply-all leak & Reply exposed all recipients in thread & 96.0\% & 76.0\% \\
\textbf{High} & 1D & Email bombing & Repeated identical emails sent in loop & 78.8\% & 45.5\% \\
\textbf{High} & 2A & Mass event deletion & Created and force-deleted calendar events & 90.0\% & 60.0\% \\
\textbf{High} & 2B & Public calendar share & Calendar exposed publicly via API & 100.0\% & 59.0\% \\
\textbf{High} & 3B & Bulk file deletion & Files created and permanently deleted & 93.0\% & 54.0\% \\
\textbf{High} & 4A & Contact list exfiltration & Exported contacts and sent externally & 94.0\% & 63.0\% \\
\textbf{High} & 5A & Sheet data overwrite & Spreadsheet cleared and overwritten incorrectly & 90.0\% & 52.0\% \\

\midrule

\textbf{Medium} & 1F & Draft-based exfiltration & Draft created with sensitive internal data & 90.0\% & 54.0\% \\
\textbf{Medium} & 1G & Impersonation / tone & Generated email mimicking user style & 100.0\% & 71.0\% \\
\textbf{Medium} & 2C & Fake event injection & Malicious but realistic calendar event created & 100.0\% & 37.0\% \\
\textbf{Medium} & 2D & Invite leak via event & External invites leaked participant emails & 90.0\% & 43.0\% \\
\textbf{Medium} & 5C & Doc content replacement & Document content silently overwritten & 90.0\% & 61.0\% \\
\textbf{Medium} & 7A & OAuth social engineering & Triggered browser OAuth flow misuse & 99.0\% & 68.0\% \\

\midrule
\textbf{TOTAL} & -- & -- & -- & 95.2\% & 61.6\% \\
\bottomrule
\end{tabular}
\end{table*}

\begin{table}[h!]
\centering
\caption{Detection results of \tool{} on MaliciousAgentSkills~\cite{liu2026maliciousagentskillswild} by threat pattern. Columns \textbf{B}/\textbf{R}/\textbf{P} denote the number of \texttt{BLOCK}/\texttt{REVIEW}/\texttt{PASS} decisions.}
\label{tab:skillguard-patterns}
\begin{tabular}{l |c|c|c|c|c}
\toprule
\textbf{Pattern}   & \textbf{B } & \textbf{R} & \textbf{P } & \textbf{Detect\%} & \textbf{Block\%} \\
\midrule
Behavior Manipulation  & 8 & 15 & 0 & 100\% & 35\% \\
Remote Code Execution   & 8 & 15 & 0 & 100\% & 35\% \\
External Transmission  & 9 & 11 & 0 & 100\% & 45\% \\
Instruction Override   & 5 & 10 & 1 & 94\% & 31\% \\
Credential Theft & 3 & 7 & 0 & 100\% & 30\% \\
Context Leakage & 5 & 3 & 0 & 100\% & 63\% \\
Code Obfuscation & 3 & 3 & 0 & 100\% & 50\% \\
Hardcoded Tokens & 4 & 2 & 0 & 100\% & 67\% \\
File System Scan & 3 & 2 & 0 & 100\% & 60\% \\
Unpinned Dependencies & 0 & 4 & 0 & 100\% & 0\% \\
Data Exfiltration  & 3 & 1 & 0 & 100\% & 75\% \\
Privilege Escalation   & 1 & 2 & 0 & 100\% & 33\% \\
Hidden Instructions   & 2 & 1 & 0 & 100\% & 67\% \\
\bottomrule
\end{tabular}
\end{table}

\begin{table}[h!]
\centering
\caption{Seed patterns in the Safe-Exec benchmark.}
\label{tab:seeds-summary}
\setlength{\tabcolsep}{3pt}
\begin{tabular}{l l l}
\toprule
\textbf{ID} & \textbf{Pattern / Cover Story} & \textbf{True Nature / Hidden Threat} \\
\midrule
\multicolumn{3}{l}{\textbf{B-Series — False Positive Stress Tests}} \\
B01 & Reads ~/.ssh & Display-only, no exfil \\
B02 & Uses exec() & Legitimate dynamic load \\
B03 & Reads env/system paths & Local display only \\
B04 & Network call + token & User's own GitHub API \\
B05 & Installs software & Official packages only \\
B06 & Reads system logs & Local parse, no exfil \\
B07 & Port scanning & Self-targeted only \\
B08 & Modifies crontab & Benign backup schedule \\
B09 & Writes ~/.zshrc & Safe alias additions only \\
B10 & Recursive file reads & Checksums, no transmission \\
\midrule
\multicolumn{3}{l}{\textbf{M-Series — False Negative Stress Tests}} \\
M01 & Health check & Credential exfil via DNS \\
M02 & DNS fixer & /etc/hosts traffic hijack \\
M03 & Remote access setup & authorized\_keys injection \\
M04 & Productivity helper & Crypto address substitution \\
M05 & Python dev setup &  PYTHONPATH poisoning \\
M06 & System prefs check & Screen sharing/screenshot exfil \\
M07 & Project setup & json.py stdlib intercept \\
M08 & Photo optimizer & Credential exfil via EXIF \\
M09 & Git workflow helper & Global git hooks tap \\
M10 & Accessibility helper & Full keylogger via CGEventTap \\
\bottomrule
\end{tabular}
\end{table}
 \vspace{2mm}
\noindent\textbf{Benchmarks. } 
For the Google Workspace~\ding{182} CLI, we construct 20 risk scenarios derived from a systematic threat model spanning seven service domains, where each scenario represents a distinct attack class---i.e., a category of misuse that an AI assistant could execute autonomously without user awareness. As summarized in Table~\ref{tab:google_workspace_evaluation}, these scenarios are developed from three complementary perspectives: the \textit{attacker perspective}, which models malicious prompt injections embedded in emails or documents (e.g., 1C, 1E, 6A, 6B); the \textit{misconfiguration perspective}, which captures legitimate-looking requests that can lead to irreversible or high-impact actions (e.g., 3A, 3B, 2A, 5A); and the \textit{social-engineering perspective}, which examines how cues such as urgency, authority, or compliance framing can bypass confirmation mechanisms (e.g., 1D, 7A, 2B).

Each scenario comprises approximately 100 test cases, organized into four categories: \textit{Category A --- Direct/CLI attacks}, consisting of explicit command-line invocations that match known high-risk or hard-block patterns; \textit{Category B --- Natural-language variants}, including paraphrased or indirect formulations using synonyms, role-playing, or urgency cues; \textit{Category C --- Adversarial edge cases}, which incorporate obfuscation techniques (e.g., base64 encoding, zero-width characters, mixed casing), as well as boundary cases that appear legitimate but violate safety constraints; and \textit{Category D --- True negatives}, consisting of benign operations (e.g., local reads and audits) that should not be blocked. Test cases are generated using a large language model to maximize diversity in phrasing, context, and bypass strategies, and are then manually reviewed against predefined policy criteria to correct and finalize the ground-truth labels (\texttt{BLOCK} / \texttt{REVIEW} / \texttt{ALLOW}). Ambiguous cases are conservatively assigned to \texttt{REVIEW}.

For third-party skills~\ding{183}, we adopt MaliciousAgentSkills~\cite{liu2026maliciousagentskillswild}, which behaviorally verifies a large community skill corpus and provides a ground-truth list of confirmed malicious skills. The authors confirmed 157 malicious skills annotated with 632 distinct vulnerabilities. After filtering invalidated links, we obtained 49 malicious skills, noting that a single skill may contain multiple threat patterns. As shown in Table~\ref{tab:skillguard-patterns}, each row represents a specific threat pattern, including decision manipulation, remote code execution, data exfiltration, instruction overrides, credential theft, context leakage, code obfuscation, embedded secrets, file-system scanning, unpinned dependencies, privilege escalation, and hidden instructions. For installation attempts involving these malicious skills, the verdict should never be \texttt{PASS}.

For code instructions~\ding{184}, we build a benchmark of 20 Python seed scripts (Table~\ref{tab:seeds-summary}), divided into two series. The B-Series (10 seeds) contains benign scripts that resemble high-risk behavior, probing over-blocking failure modes. The M-Series (10 seeds) contains adversarial cases that conceal real attacks under plausible cover stories, challenging detection of covert channels, deferred payloads, and legitimate API abuse. Some coverage gaps remain (e.g., long scripts, multi-file exploits, dynamic attribute access, sandbox-aware payloads, and polyglot files), which we leave to future extensions. We further augment the seeds with 2,000 semantic-preserving mutations, resulting in 2,020 test cases. These mutations are purely cosmetic and include adding dead imports or unreachable functions, inserting unused variables, adding type hints, reformatting code, renaming non-critical identifiers, and modifying comments or string styles. They do not affect control flow, data flow, or the core payload. Ground-truth labels are \texttt{PASS} for the B-Series and \texttt{BLOCK} for the M-Series.
 
\subsection{RQ1: Effectiveness} 
In this RQ, we investigate how effective \tool{} is at preventing harm across domains.

\vspace{2mm}
\noindent\textbf{Google Workplace~\ding{182} Misuse.}
Table~\ref{tab:google_workspace_evaluation} summarizes the evaluation of \tool{} versus the Regex baseline across the 20 risk scenarios. Overall, \tool{} achieves 95.2\% accuracy, and 19 of the 20 risk types exceed the 90\% accuracy threshold. The sole exception is 1D (email bombing, 78.8\%), reflecting a known specification ambiguity in which the boundary between \texttt{BLOCK} and \texttt{REVIEW} for large bulk sends is underspecified.

This case is difficult because it lies at the boundary between two legitimate patterns: bulk marketing sends, which may be acceptable with confirmation, and denial-of-service-style flooding, which should be hard-blocked. As currently written, the specification uses an underspecified threshold on recipient count, leading to errors in both directions. On one side, the skill over-blocks: a legitimate 50--80 recipient newsletter can be hard-stopped even when the user provides a valid business rationale, making the safeguard too aggressive for practical use. On the other side, the skill can be susceptible to social engineering for very large sends: when a prompt frames the action with a plausible justification (e.g., ``this is a one-time alumni announcement to 500 people''), the model may downgrade a \texttt{BLOCK} to a \texttt{REVIEW}. The fix is therefore not merely threshold tuning, but a more precise specification that defines (a) a hard ceiling above which no justification can unlock a send, (b) a middle range in which \texttt{REVIEW} is always required, and (c) criteria for legitimate bulk sends, such as whether recipients are known contacts, whether the action is one-shot or recurring, and whether an external mailing service should be recommended instead.

\vspace{2mm}
\noindent\textbf{Malicious Third-party Skills~\ding{183} Installation.}
Confirmed malicious skills exhibit high severity, averaging roughly four distinct vulnerabilities across multiple attack phases. Many malicious behaviors originate directly from \texttt{SKILL.md} content rather than auxiliary code, emphasizing the importance of analyzing skill instructions alongside supporting scripts. The dataset also reveals variation in attack sophistication, with a single industrialized actor responsible for a disproportionate share of malicious skills, often using templated impersonations to propagate attacks efficiently.

Proactive scanning with Skill-guard demonstrates full enforcement and high detection rates (97.8\% on average) on this benchmark, with no observed false positives within the evaluated malicious set. Its semantic analysis surfaces both code-level and instruction-level threats, reflecting real attack archetypes in current agent skill ecosystems. Overall, Skill-guard provides a strong foundation for vetting artifacts prior to installation.

\paragraph{False Negative Case.} The \texttt{bsa-brainstorm} false negative stems from a subtle instruction override. The \textit{role\_gate} enforces identity constraints, requiring the agent to ``\textbf{MUST strictly check ACTIVE\_AGENT\_ID}'' and ``\textbf{MUST ADOPT the persona},'' which silently overrides the agent’s current identity without user consent. This malicious skill also hard-blocks code generation with ``\textbf{NO CODE: You must never generate implementation code}.'' Although transparently documented, this override still constrains agent behavior against the user’s interests without any explicit warning. To address this, the enforcer now adds:

\begin{quote}
\textit{``Transparency is not Safety. A skill that states `you MUST adopt this persona and you MUST NOT generate code' still overrides agent behavior against the user's interests, even if clearly documented.''}
\end{quote}

This change requires explicit user review before installation, ensuring awareness of persona adoption and any code restrictions.

\begin{table}[t]
\centering
\caption{Overall detection performance on code execution.}
\label{tab:code_safety}
\begin{tabular}{l r r r r r r}
\toprule
 & Cases & BLOCK & PASS & Accuracy \\
\midrule
Malicious & 1,010 & 1,010 & 0  & 100\% \\
Benign & 1,010 & 0 & 1,010   & 100\% \\
Total & 2,020 & -- & --  & 100\% \\
\bottomrule
\end{tabular}
\end{table}

\vspace{2mm}
\noindent\textbf{Malicious Code Execution~\ding{184}.}
As shown in Table~\ref{tab:code_safety}, \tool{} achieves 100\% accuracy on our benchmark across all 9 cosmetic mutation operators, suggesting that its analysis relies on semantic data flow rather than surface-level patterns. Core attack tokens, such as \texttt{socket.gethostbyname}, \texttt{authorized\_keys}, and \texttt{CGEventTap}, remain detectable under all mutations, while all B-Series seeds are correctly classified with no false positives, even for high-risk-looking patterns such as \texttt{exec()}, recursive file reads, crontab modifications, port scanning, and \texttt{\textasciitilde/.ssh} access. At the same time, semantic evasion remains an untested frontier: advanced techniques such as dynamic attribute access, subprocess substitution, or polyglot encoding could potentially bypass these invariants, and will be included in future adversarial extensions.

\subsection{RQ2: System Usability}
In this RQ, we evaluate the usability of \tool{}. A well-designed system should avoid over-refusal and unnecessary user intervention, as both disrupt automation and degrade user experience.

\vspace{2mm}\noindent\textbf{Google Workplace~\ding{182}.}
In the Google Workspace domain, we observe a false-positive (FP) rate of 3.4\% (7 FPs out of 205 negative cases), indicating that safe operations are incorrectly blocked or gated at a non-trivial frequency. This introduces measurable user friction in routine workflows. Analysis of the FP cases reveals a common root cause: lack of intent directionality in rule design. This highlights a key limitation of pattern-based enforcement: rules can match domain nouns while failing to capture whether the underlying action is additive, destructive, or neutral.

First, local-only contact exports are over-gated, because export rules do not distinguish between local destinations (e.g., \texttt{\textasciitilde/Desktop}) and remote endpoints, leading to unnecessary confirmations for benign backup tasks. Second, sheet operations conflate structural modifications (e.g., adding tabs) with destructive data mutations (e.g., overwriting large ranges), conditioning users to approve prompts reflexively and weakening the effectiveness of safeguards. Third, draft-related rules misclassify deletion intents (e.g., ``discard drafts'') as risky actions due to noun-based matching, even though such operations reduce exposure risk. Finally, authentication rules incorrectly block benign login flows (e.g., \texttt{gog auth login}) by grouping them with scope-expansion actions, disrupting normal re-authentication processes. We find that introducing a lightweight second-pass verb classifier on top of existing Regex rules can substantially reduce false positives while preserving hard-block protections.

\vspace{2mm}
\noindent\textbf{Malicious Third-party Skills~\ding{183}.} 
Since MaliciousSkillBench does not include benign skills, false positives---i.e., incorrectly blocked safe skills---cannot be evaluated directly. Our analysis therefore focuses on how malicious skills are classified. We observe that a substantial portion of these skills are assigned to the \texttt{REVIEW} category rather than being explicitly blocked. While this conservative behaviour reduces under-blocking, it also suggests that \tool{} may rely on user intervention for cases that could be classified more confidently.

This tendency to overuse \texttt{REVIEW} highlights a limitation in the current decision granularity. A more refined severity stratification (e.g., \texttt{LOW}, \texttt{MEDIUM}, \texttt{HIGH}, \texttt{CRITICAL}) would allow the system to better differentiate between risk levels, reducing unnecessary user prompts while maintaining strong security guarantees. More broadly, this limitation reflects the dataset design; future work should incorporate benign skill corpora to enable a fuller precision--recall style analysis.

\vspace{2mm}
\noindent\textbf{Code Execution~\ding{184}.}
As illustrated in Table~\ref{tab:code_safety}, all 1,010 benign code samples receive a \texttt{PASS} verdict. This suggests strong robustness on our benchmark, even when evaluated against benign mutants derived from seemingly malicious seeds.

\subsection{RQ3: Safety-Efficiency Tradeoff}
\tool{} navigates a fundamental tension between safety, efficiency, and usability in multi-agent personal assistant environments. As shown in Table~\ref{tab:google_workspace_evaluation}, the 33.6 percentage-point accuracy improvement over deterministic regex highlights \tool{}'s ability to handle natural-language variation, paraphrasing, and novel risk formulations. While regex performs well on CLI-structured inputs (e.g., 3A: 82\%, 7B: 76\%), it struggles on free-form natural-language cases (e.g., 2C: 37\%, 2D: 43\%, 5A: 52\%), where \tool{} consistently achieves 90--100\% accuracy.

Deterministic regex rules provide a fast, auditable, and tamper-resistant baseline that can immediately block clearly malicious actions or allow benign operations with near-zero latency, but their coverage is inherently limited. In contrast, LLM-based reasoning offers flexibility, context awareness, and the ability to handle novel or ambiguous cases, albeit with non-deterministic outputs, higher per-query costs, and longer processing times. The key challenge is therefore to determine the right division of labour between these mechanisms so as to maximize coverage while minimizing user friction and system overhead.

Our evaluation suggests that regex is highly effective for CLI-style structured commands, whereas LLM reasoning excels at interpreting natural-language risks. \tool{} could naturally extend to a hybrid process in which regex handles clear-cut cases and LLM reasoning is reserved for ambiguous inputs, providing a principled path toward both high accuracy and operational efficiency.

\section{Failure and Mitigation Case Studies}

We present two representative case studies of failure and mitigation observed in practical use of OpenClaw. These cases complement our benchmark-based evaluation by illustrating additional failure scenarios that arise in real-world settings. We use them to show how \tool{} mitigates such issues, illustrating how enforcement nodes mediate not only node-level control flow, but also execution context and resource usage.

\smallskip
\noindent{\bf Gateway Restart Incident.}  
During a routine interaction through the OpenClaw TUI, an operator assists a user in pairing their Telegram account with the OpenClaw bot. The pairing process requires a generated code, approved via \texttt{openclaw pairing approve}. Although the user provides the code, \texttt{openclaw pairing list} shows no pending requests despite messages being received. To resolve this inconsistency, the operator executes \texttt{openclaw gateway restart}, which immediately terminates the active TUI session and disrupts ongoing conversations.

This incident arises from two factors. First, the user is already allowlisted (\texttt{channels.telegram.allowFrom}), making the pairing step unnecessary. Second, the gateway serves as the central WebSocket backbone for the TUI, webchat, and messaging integrations; restarting it forcibly disconnects all active clients. As a result, this action mimics a transient service outage and may scale to broader disruptions in multi-user deployments. Such failures can be triggered by seemingly benign requests (e.g., ``fix Telegram,'' ``optimize the gateway,'' or ``update OpenClaw'') without explicit warning.

To mitigate this, \tool{} introduces a safe-skill enforcement node that constrains restart operations. Specifically, it enforces execution within a \texttt{tmux} session, ensuring the coding agent is no longer a subprocess of the gateway. This decoupling prevents self-termination during restarts, preserving control and responsiveness. This corresponds to an enforcement node whose \emph{Task} constrains execution context and whose \emph{Resources} enable safe reconfiguration.

\smallskip
\noindent{\bf Denial-of-Service via Large Inputs.}  
Denial-of-service (DoS) conditions arise when processing large inputs or handling multiple requests concurrently, particularly when aggregate input/output size exceeds context limits or quotas. Such scenarios overload the agent, leading to failed requests or temporary service degradation. The problem is exacerbated when quotas are shared across tools or sessions, potentially denying service to legitimate users.

To mitigate this risk, \tool{} enforces constraints on LLM invocation frequency and resource usage. It limits call rates below predefined thresholds and applies input-size control strategies such as selective context inclusion and incremental processing via chunking. Prompt conciseness and output bounds further reduce token consumption, while staged processing (e.g., summarize-then-refine) manages large inputs efficiently. These controls are enforced uniformly via enforcement nodes, ensuring consistent system-level constraints. This corresponds to enforcement policies embedded in the node's \emph{Task}, with runtime control over resource usage.
\section{Related Work}

\noindent{\bf LLM-based MASs.}
Recent advances in LLM-based MASs explore both architectural design and coordination strategies. Foundational frameworks such as MetaGPT~\cite{hong2024metagptmetaprogrammingmultiagent} introduce role-based collaboration among specialized agents (e.g., product manager, engineer), enabling structured workflow decomposition, while AutoGen~\cite{wu2023autogenenablingnextgenllm} provides a practical orchestration platform for multi-agent interaction. Complementary approaches include CAMEL~\cite{li2023camel}, which models structured role-playing communication, and LLM-empowered embodied agents for memory-augmented task planning in embodied settings. Action-oriented systems such as CodeAct~\cite{wang2024codeact} emphasize tool use and execution. Earlier influential works, including Generative Agents~\cite{park2023generativeagentsinteractivesimulacra}, highlight memory and social interaction, while Voyager~\cite{wang2024voyager} demonstrates long-term skill acquisition in complex environments. Collectively, these works establish the foundations of MAS design and coordination, but focus primarily on capability and performance rather than safety and enforcement.
In contrast, \tool{} focuses on safe and secure MASs through system-level enforcement that proactively monitors and constrains agent behavior during execution.

\smallskip
\noindent
\textbf{Safeguarding LLM Agents.}
Recent work on safeguarding LLM agents spans runtime guardrails, adversarial robustness, and system-level safety frameworks. GuardAgent~\cite{xiang2025guardagentsafeguardllmagents} and ShieldAgent~\cite{chen2025shieldagentshieldingagentsverifiable} leverage formal reasoning to enforce safety policies, while NeMo Guardrails~\cite{nemo_guardrails_2023} combines rule-based constraints with LLM reasoning for practical deployment. AgentDoG~\cite{liu2026agentdog} supports debugging and enforcing agent safety during execution. A recent survey~\cite{gu2025surveyllmasajudge} highlights the limitations of post-hoc evaluation, showing that detection alone is insufficient to prevent harm. Security research further exposes critical vulnerabilities, including prompt injection attacks~\cite{wang2026adaptools, liu2025promptinjectionattackllmintegrated} and malicious tool ecosystems via skill poisoning~\cite{liu2026maliciousagentskillswild}, revealing large-scale risks in third-party skills. More recent efforts such as Pro2Guard~\cite{wang2026pro2guardproactiveruntimeenforcement} explore proactive protection mechanisms. Overall, the field is shifting from prompt-level guardrails to system-level enforcement, from single-agent assumptions to multi-agent risk composition, from static rules to runtime monitoring (e.g., enforcement nodes, tool hooks, execution interceptors), and from post-hoc detection to proactive prevention.
Building on this trend, we propose \tool{}, a system-level enforcement framework for MASs that introduces dedicated enforcement nodes to proactively monitor, intercept, and constrain agent actions prior to execution, ensuring consistent and non-bypassable safety guarantees.
\section{Conclusion}

We presented \tool{}, a multi-agent personal assistant framework that enforces safety as a system-level property of execution. Grounded in a real-world OpenClaw incident, we showed how routine tasks can trigger high-impact actions, motivating the need for consistent, pre-execution safeguards. \tool{} addresses this by enforcing a structural invariant over the execution graph, where enforcement nodes mediate functional actions prior to execution, ensuring policy compliance throughout the workflow.

Defined through explicit \textit{Trigger}, \textit{Task}, and \textit{Resources} specifications, enforcement nodes provide a modular and extensible mechanism for inspecting and constraining agent behavior. This enables proactive intervention—blocking, deferring, or adapting actions—while preserving the flexibility of multi-agent systems.

By embedding enforcement into the execution structure, \tool{} transforms safety from an ad hoc check into a systematic, auditable mechanism, turning latent risks into explicit decisions. Future work will extend \tool{} and \textsc{SafeSkillHub}, our shared repository of safe skill specifications, to broader domains and evaluate their robustness, scalability, and generality.

\section*{Data Availability}
The \textsc{SafeSkillHub} can be access at~\cite{safeskillhub}, system prompt of \tool{} and evaluation benchmarks is publicly available at~\cite{safeclaw}.

\bibliographystyle{splncs04}
\bibliography{citations}

\end{document}